# Crystallographic and Electronic Phase Changes in $TiTe_2$ via Atmospheric and Electron Beam Exposure

Bishal Pokhrel[1,*], Joel Quarnstrom[1], Saraswati Shrestha[1], Halle Helfrich[1,2], Elena Echeverria[1,3], David N. McIlroy[1], and Andrew J. Yost[1,4]

[1]*Department of Physics, Oklahoma State University, Stillwater, OK, 74078*

[2]*Department of Physics, Pittsburg State University, Pittsburg, KS, 66762*

[3]*The Center for Bright Beams, Cornell University, Ithaca, NY, 14853*

[4]*Scienta-Omicron Inc, Denver, CO, USA*

**Corresponding author: bishal.pokhrel10@okstate.edu*

*ORCID: 0009-0004-8378-3254*

## Abstract

In this study, we examine the surface sensitivity of a transition metal dichalcogenide ($TiTe_2$) grown using the Chemical Vapor Transport (CVT) technique at high pressure and study the surface changes in the sample upon exposure to air as well as its crystallographic properties upon e-beam exposure. We examine the local density of States (LDOS) of the mechanically exfoliated sample surface before and after exposure using Scanning Tunneling Microscopy/Spectroscopy (STM/STS) and find a metal-to-semiconductor transition at the surface. The STM analysis shows a clear change in surface roughness which indicates the formation of an adlayer on the surface of the sample. The crystal structure was examined using X-ray Diffraction (XRD) and Transmission Electron Microscopy (TEM), indicating a phase transition from a single-crystalline hexagonal phase to a polycrystalline state via a distorted monoclinic phase. Furthermore, the presence of a possible superlattice suggests non-stoichiometric $Ti_xTe_y$ at the surface of the exposed layer.

## I. Introduction

Transition Metal Dichalcogenides (TMDs) of the form $MX_2$, where M represents a transition metal and X represents chalcogen elements, have been studied extensively due to their interesting properties like high carrier mobility [1,2], direct as well as indirect bandgap, room-temperature ferromagnetism [3,4], charge density waves [5,6], among others. The physical, chemical and electronic properties of these 2D materials are versatile and material dependent, i.e. by just changing the transition metal or the chalcogen atom or both, significantly different properties can be extracted from the material. For example: $NbS_2$ is a metal [7], $MoS_2$ is a semiconductor [8], $HfS_2$ is a semi-insulator [9], while $NbSe_2$ is a superconductor [10,11]. This happens due to the presence of unfilled d-orbitals in these transition metals. Quantum confinement in layered d-electron materials, such as $MoS_2$, has been shown to result in a transition from an indirect to a direct bandgap when thinned to a monolayer [12,13]. The effect of going from bulk to 2D has been shown in materials like GeP nanosheets where the bandgap changes from 0.9 eV to 2.3 eV, with mobility and on/off ratio depending on the thickness of the material [14]. $TiTe_2$, like other dichalcogenides, has been shown to have a 1T-$CdI_2$ like structure, characterized by hexagonal sheets of Ti sandwiched between sheets of Te, forming X-M-X layers separated by a van der Waals gap [15,16]. Due to their unique nature, TMDs have applications in various branches of science. They possess high surface area and good electronic conductivity, which make them attractive as electrodes for energy storage devices [17] and high electron mobility and good on/off ratio which are essential for electronic devices like field-effect transistors (FETs) and logic gates, with materials like $MoS_2$ and $WSe_2$ already studied for their potential use in optoelectronics applications such as solar cells, photodetectors, and light-emitting diodes (LEDs)[18-21].

2D materials are at the forefront of nanoscale science, and their surface sensitivity has provided researchers a new parameter to experiment with. The surface sensitivity of TMDs varies depending on the growth conditions and the transition metal/chalcogen combination used. TMDs like $MoS_2$, $WSe_2$ and $TiTe_2$ in bulk form grown with conventional methods are inert to ambient air. MBE grown $SnSe_2$ however has been shown to be unstable in air, while single-layer $MoS_2$ has also been shown to be susceptible to oxidation [22-24]. With a variation in chalcogen concentration, we can grow a different class of materials, called the transition metal trichalcogenides (TMTs). They typically have a quasi-1D structure because of strong in-plane bonds in a particular direction. Of the class of TMTs, titanium tritelluride ($TiTe_3$) is particularly intriguing; the synthesis of which has only been conducted recently in confined multiwall carbon nanotubes [25,26] since it is unstable in its bulk form. Hence experiments on the material, apart from in-situ studies, cannot be conducted without oxidizing it. While the synthesis of monochalcogenide (TiTe), dichalcogenide ($TiTe_2$), and sesquichalcogenide ($Ti_2Te_3$) has been shown [27], with all of them possessing hexagonal symmetry, as of yet, the growth of free-standing $TiTe_3$ has not been reported, with the only reported synthesis being in reaction vessels like the carbon nanotubes.

Various growth techniques like mechanical exfoliation, chemical vapor deposition [28,29], molecular beam epitaxy [30], and different chemical methods have been used to grow TMDs so far but the high-pressure growth and investigation of these TMDs, especially at the nanoscale, is relatively unexplored. In this paper, we seek to tweak the growth parameters from a normal CVT growth process and investigate the resulting sample to characterize its surface response. The local density of states at the surface before and after the contamination has been characterized using STM/STS, and the bonding environment of the corresponding elements have been examined using X-ray photoemission spectroscopy (XPS) to understand the effects of air exposure on surface dynamics. We also examine the crystal structure of the sample using X-ray Diffraction (XRD) and the change in crystal symmetry through Transmission Electron Microscopy (TEM).

## II. Sample Growth

Bulk $TiTe_2$ was grown using the Chemical vapor transport (CVT) technique. The sample preparation and CVT growth was executed in a sealed glass ampule. Before the use of the ampule, it was rinsed with deionized water and dried to reduce the amount of impurities in the sample. For bulk synthesis of the material, mixture of 0.0528 g Ti foil and 0.2071 g Te powder was evacuated in a glass ampule roughly 12 cm in length, followed by back filling with Nitrogen (to create an inert and controlled environment), finally pumping down to pressures between 5000 mTorr to 500 mTorr such that pressure inside the ampule is above 1 atm during the heating procedures. Note that the starting concentration of Ti:Te was not 1:2. The ampule was vacuum/pressure sealed using a hydroxy torch (to prevent any carbonaceous contamination from fuel source), and heated to 900-1000 °C at a rate of 10 °C/min in a tube furnace for 5 days and allowed to cool to room temperature naturally. This constitutes one heating cycle. The heating cycle was repeated 3-5 times until the Te powder sublimates and reacts with the Ti foil completely forming the desired compound. The resulting material exhibits a highly reflective mirror-like surface with a silver color (Fig. 1(a)), atypical of conventionally grown $TiTe_2$, which is grayish-black in appearance. It begins turning black in patches across the surface when exposed to atmosphere for extended periods of time, t > 10 min. Manipulating the pressure and stoichiometry can also yield nanowhisker form of the same material akin to the quasi-1D structure of the TMTs (Fig. 1(b)). The experiments described in this article were done on the sheets/flakes of $TiTe_2$. The nanowhiskers will be examined in future experiments.

## III. Experimental Methods

X-ray diffraction measurements were performed with a Rigaku high-definition diffractometer using a Cu-Kα source (λ = 1.5418 Å). The tube voltage was set to 40 kV and the current to 15 mA. The measurements were taken at room temperature in standard atmospheric conditions.
Transmission electron microscopy (TEM) and selected area electron diffraction (SAED) were performed on a JEOL JEM-2100 scanning transmission electron microscope at an accelerating voltage of 190 kV.
X-ray photoemission (XPS) was performed in a UHV chamber that maintains a base pressure better than $8.0 \times 10^{-10}$ mbar. XPS spectra were acquired using an Mg-Kα X-ray source (Physical Electronics XR 04-548), at 400 W and incident angle of 54.7°. The kinetic energy of the photoelectrons was recorded with an Omicron EA 125 hemispherical electron energy analyzer with a resolution of 0.02 eV. All measurements were performed at room temperature.
Scanning tunneling microscopy and spectroscopy measurements were performed with a Scienta Omicron Gen III LT-STM using an atomically sharp Pt-Ir tip in an ultra-high vacuum (UHV) chamber with a base pressure better than $1.0 \times 10^{-11}$ mbar. Measurements for all samples were performed at 298 K. All topography images and dI/dV measurements are made with the STM tunneling current set point, $I_t$ = 1 nA, and tunneling bias set-point, $V_{gap}$ = 1.1 V, unless mentioned otherwise.
The bulk Ti-Te crystals were exfoliated using an adhesive tape. As the material has weak van der Waals interaction, peeling off the tape removes the major portion of the crystal and few-layer to multi-layer flakes can be extracted and then transferred onto the substrate.
The STM/S measurement was done on the mono to few-layer material after exfoliating inside the vacuum chamber and then returned to the load lock after which it is exposed to the atmosphere for time t > 10 min. The sample was returned to the STM and the same measurements were done on the exposed material. This process was repeated several times to ensure the results were repeatable. Pressure change is from $1x10^{-10}$ mbar to 1013.18 mbar (i.e. ~ 1 atm).

## IV. Results and Discussion

### *i. Crystal Structure*

Once the sample was exposed to air in the laboratory setting, we saw some signs of changes in the surface as the silvery-white surface started turning grayish-black which is what conventional 1T-TiTe2 looks like. $TiTe_2$ is also known to be very stable in air. XRD measurements were performed on the sample to understand its crystal structure. The X-ray Diffraction (XRD) pattern shown in Fig. 2 shows strong (00n) peaks indicating the growth of single crystals along the c-axis, consistent with the diffraction patterns reported for 1T-$TiTe_2$. It has a trigonal crystal symmetry and space group P-3m1 with lattice parameters in close match with those reported for 1T-TiTe2 i.e. a = b = 3.774 Å, c = 6.415 Å and α = β = 90°, $\gamma$ = 120° [5,31,32]. The $2\theta$ angles corresponding to the reported peaks are 14.1°, 27.9°, 42.2°, and 57.1° respectively. Diffraction pattern using SAED (Fig. 3(a)) also confirmed the highly ordered hexagonal crystal structure along the [001] axis, and the single crystal nature of the sample, with distinct dots representing each diffracting plane. This confirms that the grown material is indeed $TiTe_2$. The surface changes seen will be explained in a later section. For now, its response to radiation exposure will be studied.

### *ii. E-beam Exposure*

Bulk $TiTe_2$ was sonicated with ethanol, and the prepared sample was exposed to transmitted beam from TEM electron gun at 190 keV. After exposure to radiation, we saw changes in the sample after around 1-2 min, suggesting the sample interacts with the e-beam. The data presented in this segment will show a comparison of sample's behavior before and after long-term exposure (> 2 min) to the radiation. Upon first inspection, significant mass loss was observed in the sample as shown in Fig. 4(a, b) and similar changes were observed in the lattice fringes. The inter-layer separation before exposure is around 0.35 nm which is consistent throughout the sample (Fig. 4(c)). After exposure, the inter-layer spacing varies in the range 0.35 nm to 0.47 nm (Fig. 4(d)) along with the formation of distinct domains visible on the HR-TEM image. This corresponds to the beam induced artifacts/disorder on the sample, suggesting the sample undergoes structural transformation, leading to the formation of grain boundaries and domains with different chemical composition. We attribute it to the solvent used in the sample preparation process, ethanol. As the high energy electron beam probes into the sample, it breaks the ordered nature of the sample and the -OH groups from the solvent, which are present on the surface, recombine with the sample potentially forming $TiO_2$, $Te(OH)_6$, or some combination thereof. This explains the formation of domains in the crystal with these new elemental contributions changing the spacing between the lattice planes.
This result is also apparent from the SAED spectra of the sample. Fig. 3(a) shows the perfectly ordered trigonal/hexagonal crystal structure with minimal exposure to e-beam which transitions to a distorted monoclinic-like structure (Fig. 3(b)) upon ~ 1-2 min of exposure. At this point, one point to note is that $TiTe_3$ is predicted to crystallize in a monoclinic structure [25,33], but this phase potentially being a trichalcogenide is beyond the scope of this study. Meanwhile, $TiTe_2$ has been shown to undergo phase change from hexagonal to monoclinic at a pressure of around 5.4 GPa [34] and 8 GPa [35]. These studies subject $TiTe_2$ to such high pressures after growing the crystal through conventional methods, which is clearly different from our method. But it gives a relevant insight that $TiTe_2$ subjected to high pressures is inclined to undergo structural transitions. Irradiating with the high energy beam for more than 2 min turns the crystal polycrystalline (Fig. 3(c)). One can deduce that further exposure to the beam will make the sample amorphous, losing all its crystalline properties. Radiation induced damage in the crystal structure has also been studied in graphene [36], where disorder starts at 5-20 keV with continuing evolution to nanocrystalline and amorphous phases as the accelerating voltage is increased. Hybrid perovskites like $CH_3NH_3PbI_3$ have been shown to undergo structural and compositional transformation upon exposure to electron beam [37].
The inset in Fig. 3(b) shows multiple lobes at each diffraction spot reminiscent of a super periodic structure, confirmed as not relating to the aberration that the TEM setup has. This suggests the breaking down of the crystal into different diffraction planes as it interacts with the beam and starts forming grain boundaries that results in domain formation as was suggested earlier. Although TEM is not a surface sensitive technique, some literature exists on the effect of electron diffraction on sensitive surfaces. Joyner et al. [38] observed multiple diffraction spots through Low Energy Electron Diffraction (LEED) on Pb (100) and Pb (110) surfaces which were attributed to the oxide and sulphide formation at the surface. This is not surprising since 2D materials are highly susceptible to interacting with their surroundings with materials like Black Phosphorus (BP), Black Arsenic and other ternary 2D materials like $Cr_2Ge_2Te_6$ showing poor air stability [39].
Similar conclusions can be drawn in the case of the high-pressure grown $TiTe_2$. The observation of a super periodic structure, coupled with the instability of the material when exposed to ambient air point towards the surface/electronic reconstruction of the sample upon exposure - most likely due to the surface contamination due to the hydroxyl species attaching from the solvent and/or air. This is in stark contrast to that of $TiTe_2$ crystals grown at normal pressures, which cleave at the van der Waals gap, resulting in stable and non-reactive surfaces without any detectable degradation or contamination for several days.

### *iii. Changes in Chemical Composition*

To evaluate the changes in chemical content on the sample because of e-beam exposure, Energy-dispersive X-ray spectroscopy was taken from the same TEM setup. EDS spectra of the sample before and after long-term exposure are shown in Fig. 5. For reference, the Cu peaks in the spectra are due to the Copper grid used in the sample preparation stage, and the Silicon peak is due to the internal fluorescence peak due to the detector. Initially, titanium (Ti) and tellurium are the only significant peaks like we would expect, as the sample is initially in an ordered crystalline form. After exposure, however, the relative intensities of Ti and Te reduce significantly in favor of Oxygen (O). This supports our hypothesis of oxide and hydroxide formation after consistent e-beam exposure. One must note that lighter elements like hydrogen (H) which may be present on the sample could not be detected as the EDS setup cannot detect elements below carbon (C). This calls for a more detailed study of the sample surface and the possible reconstruction, which will be analyzed in the next section.

### *iv. Bonding environment before and after exposure to air*

Returning to the effects of atmospheric contamination observed earlier; to understand the surface dynamics of the sample, XPS study of the surface was done before and after long-term air exposure. Survey scans done on the sample (Fig. 6) show distinct peaks owing to contributions from Ti, Te, O and C core levels at various binding energies. New peaks can be seen in the air exposed sample above BE of around 750 eV, representing Auger transitions in Ti and Te core levels. So, Auger electron emission is playing a significant role in the relaxation process of the sample after X-ray photoionization. The survey scan also does not have Nitrogen core level features i.e. the sample is not a nitride-based material (MXene) [40]. This suggests Nitrogen was not incorporated into the crystal during the evacuation process. We can also see that the Ti peaks are not as prominent as the Te peaks due to the atomic sensitivity factor (ASF) of Ti and Te, which are 2.001 and 5.705 respectively.
In order to assess the surface composition and bonding environment, high resolution scans of each atom's core levels were performed. The high-resolution scan of C 1s (Fig. 7(a)) shows multiple peaks from 284.5 eV to 292.1 eV. We attribute these peaks to the carbon sticky tape used to mount the sample and the adventitious carbon from exposure to the laboratory environment upon retrieval from the quartz ampule. O 1s, on the other hand, shows satellite peaks around 530 – 533 eV (Fig. 7(b)). This can be attributed to the formation of metal oxides and hydroxides at the surface. The peak at 530.50 eV is due to $TiO_2$, the presence of which will be clearer after examining the high-resolution spectra of Ti 2p core level. The Auger lines that were seen indicate that when the relaxation process happens, oxygen bonds with the transition metals at the surface, which along with some charging effects on the sample surface can be attributed to the shift in the binding energy of the O 1s characteristic peak, as well as the peak splitting observed at 535.76 eV and 537.51 eV.
We compared the core level peaks of Ti 2p and Te 3d before and after exposure to air for t > 10 min. For titanium, the characteristic Ti $2p_{3/2}$ and Ti $2p_{1/2}$ peaks were seen at 455 eV and 461.07 eV respectively with satellite peaks at 457.9 eV and 463.6 eV corresponding to Ti in $TiO_2$[41] (Fig. 8(a)). We see these oxide peaks because the sample was exposed to air for a few minutes while preparing the sample for XPS mounting. After that, the sample was left in air for more than 10 minutes and then mounted onto the XPS stage. This results in a clear reduction in peak intensity of Ti core levels in favor of excess $TiO_2$ formation (Fig. 8(c)). For tellurium, Te $3d_{5/2}$ and $3d_{3/2}$ were seen at 572.14 eV and 582.53 eV respectively with satellite peaks at 575.64 eV and 586.41 eV indicative of $TeO_2$/$Te(OH)_6$ formation (Fig. 8(b)). After exposure, Te-O/Te-OH peaks do increase in intensity (Fig. 8(d)), but not to the extent where they dwarf the core Te 3d peaks, unlike Ti. So, during the surface reconstruction, Oxygen favors bonding with titanium more so than tellurium, resulting in a $TiO_2$ rich surface for the exposed sample. This is interesting since we expect the mechanically exfoliated $TiTe_2$ to terminate in Te more often than Ti. We believe it is because $TiO_2$ is more thermodynamically stable than $TeO_2$. So, we can conclude that the surface is not pure $TiTe_2$ but has oxides and hydroxides of transition metals present.

### ***v. Surface topography and density of states before and after exposure to air***

Bulk crystals were mechanically exfoliated to a few-layer limit and STM topography data of the sample were collected before and after the exposure to air through the load lock. For a freshly exfoliated surface, adhesive tape was attached at the end of the sample loading arm, and the exfoliation was done in-situ inside the STM main chamber. A clear decrease in RMS roughness was observed from 27.23 nm when freshly exfoliated to 12.52 nm after exposure indicating an adlayer/s formation at the crystal surface.
The corresponding STS profile of the sample is shown in the I-V curve (Fig. 9) where we observe that bulk $TiTe_2$ single crystals exhibit a metallic density of states behavior with flat band conditions, when freshly exfoliated, transitioning to semiconducting upon exposure to the atmosphere. This suggests an n-type behavior in vacuum and p-type after exposure, due to p-doping effects as a result of surface contamination. The semiconducting phase shows a small bandgap of around 8.22 meV. The air-exposed sample was exfoliated in vacuum again where the metallic behavior resurfaced i.e. the material goes from metallic to semiconducting, then back to metallic upon repeated exfoliation of the reconstructed surface. Thus, we can safely say that the reconstruction happens only at the surface, preserving the crystallinity in the bulk sample underneath the surface. This indicates that our material is highly chemically sensitive to molecular species present in our atmosphere like –OH, -$H_2O$, $O_2$, etc. These molecules react with the surface of the sample and create adlayer/s on top, decreasing the conductivity of the material that results in semiconducting density of states as seen in the STS profile. These adlayer(s) also contribute to the super periodic structure seen previously in the SAED pattern, where we observe additional diffraction spots corresponding to the diffraction planes of Ti and Te oxides and hydroxides. The disorder caused by rapid surface reconstruction also prevents the atomic resolution imaging in STM.
At room temperature, $TiTe_2$ is a semi-metal with a bandgap of -0.8 eV [5] with conventional metallic behavior at temperatures above 5 K, and a semiconducting phase expected at low temperatures, typically < 1.5 K due to the formation of a charge density wave with a stable 2×2 reconstruction in single layer at $T_{CDW}$ = 92 K [23,42]. $TiTe_2$ grown at normal pressure also does not show any signs of electronic or structural phase transitions [43]. Likewise, $TiO_2$ is a semiconductor with a bandgap of around 3.2 eV [44] and tellurium oxide films have bandgap of 3.5 - 3.75 eV [45]. As the bandgap in the semiconducting phase of our sample was considerably lower, we conclude that the surface is not oxidized uniformly, and there could be other molecules attached to the surface. Therefore, the sample surface is no longer a dichalcogenide after air exposure, rather turns into some form of $Ti_xTe_y$ at the surface due to contamination and the subsequent surface reconstruction at room temperature.

## V. Summary

In this study, we synthesized and characterized $TiTe_2$ (001) grown at higher pressure, unlike conventional vapor transport techniques. We observed the surface sensitivity of the sample upon exposure to solvents (chemical sensitivity) and air with changes in its crystallographic phases and surface structure. We also observed metallic vs semiconducting behavior in the local density of states for mechanically exfoliated $TiTe_2$ (001). This suggests that the surface rapidly decomposes when exposed to atmosphere, against convention, forming adlayer/s at the surface. This is reminiscent of the formation of an unstable $Ti_xTe_y$ surface on the bulk $TiTe_2$ due to -OH, $O_2$, -H, $H_2O$ ad-atoms.
From the results of this experiment, we can identify potential future applications related to this material. Firstly, because of its chemical sensitivity and high specific surface area (being a 2D material), it could be used for chemical sensing device applications where it can be used to host specific molecules that interact with target analytes. This makes it a potential candidate for gas sensors as an alternative to metal-oxide-based sensors [46,47]. Furthermore, we believe it will have plenty of applications in electronic and optoelectronic devices like transistors, diodes and memristors [48]. Such devices have already been fabricated using a variety of TMDs as discussed in the introduction section. Owing to its transition between metallic and semiconducting phases, a logic gate can be fashioned with the material that acts as an on/off switch. These logic gate applications remain an area for future exploration.

## Author Contributions

A.J.Y. designed the experiment and supervised the execution, J.Q. grew the crystals and performed XRD measurements, E.E. and D.N.M. performed the XPS measurements, B.P. performed the TEM, XRD, SAED measurements. B.P., S.S., and H.H. performed the STM and STS measurements. B.P. performed the analysis of the associated data. B.P. and A.J.Y. wrote and edited the manuscript. All authors contributed to the writing and editing of the manuscript.

## Competing Interests

A.J.Y. is currently employed by Scienta Omicron Inc. The measurements reported here were performed at Oklahoma State University prior to this employment. The remaining authors declare no competing interests.

## Acknowledgements

A.J.Y. and B.P. would like to acknowledge the support provided by Oklahoma State University in the form of startup funding which partially supported this work. D.N.M. would like to acknowledge funding provided by the Office of Naval Research (No. N00014-20-1-2433) and the Air Force Office of Scientific Research (No. FA9550-21-1-0456).

## Figures

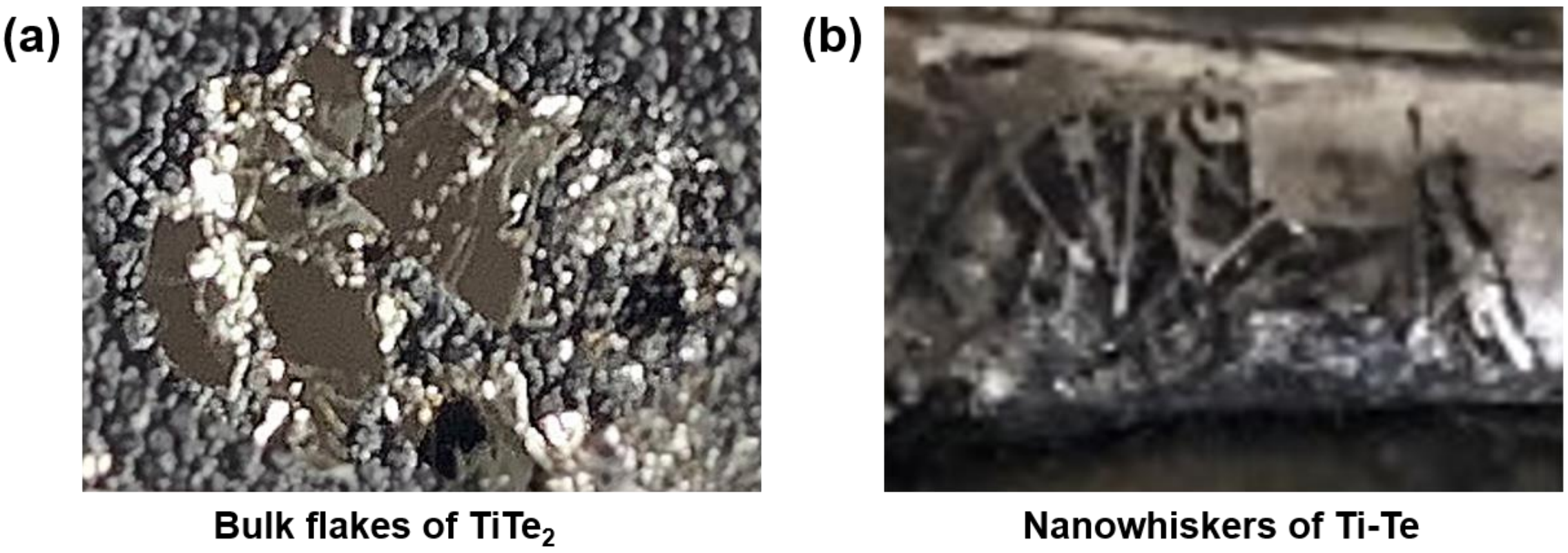

**Fig. 1:** (a) Bulk Sheets of $TiTe_2$ silver-white in color with some spots turning blackish after exposure to air. (b) Nanowhiskers grown from the same technique with different pressure.

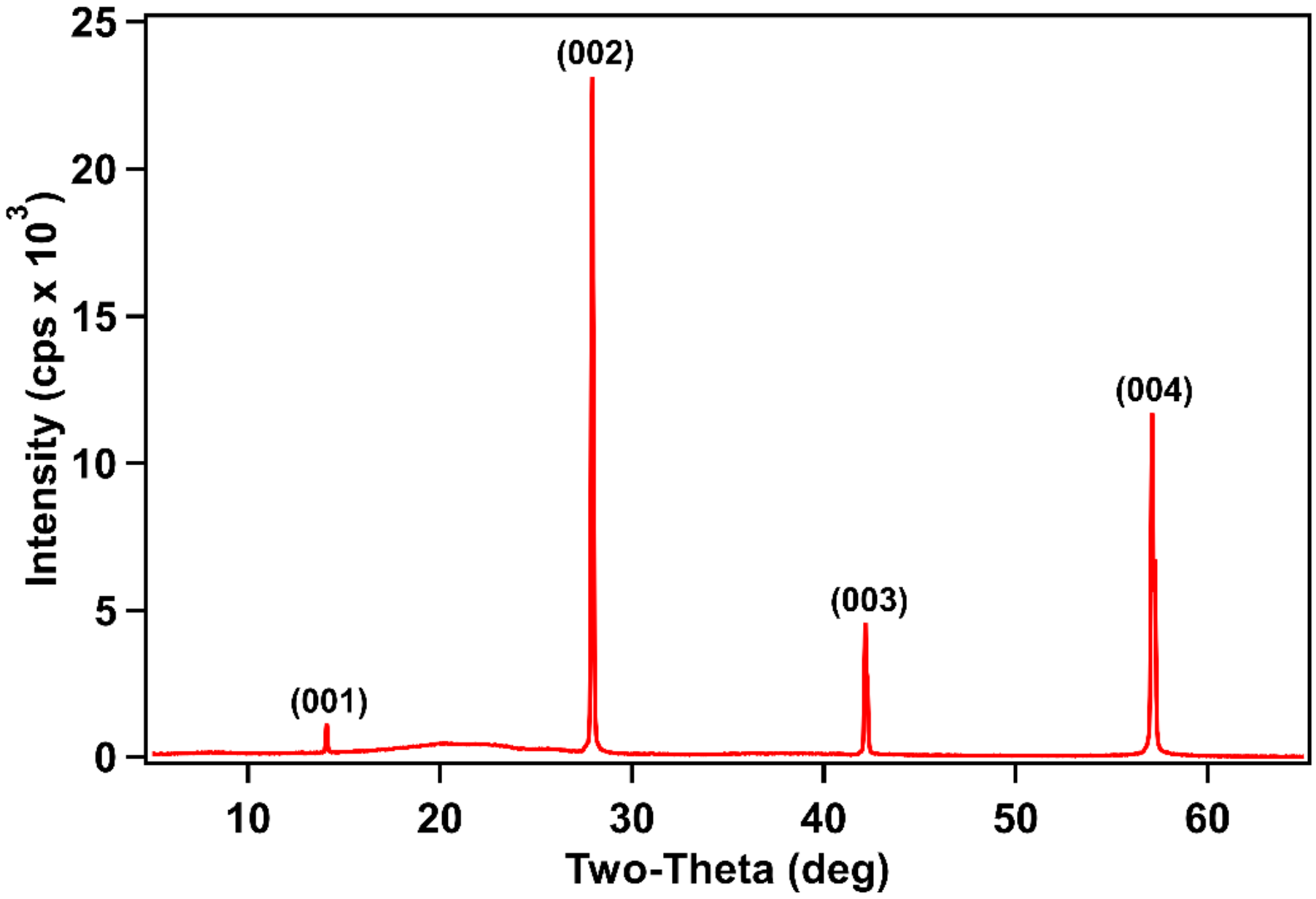

**Fig. 2:** XRD pattern of $TiTe_2$ showing distinct (00n) peaks, indicating growth of ordered single crystals.

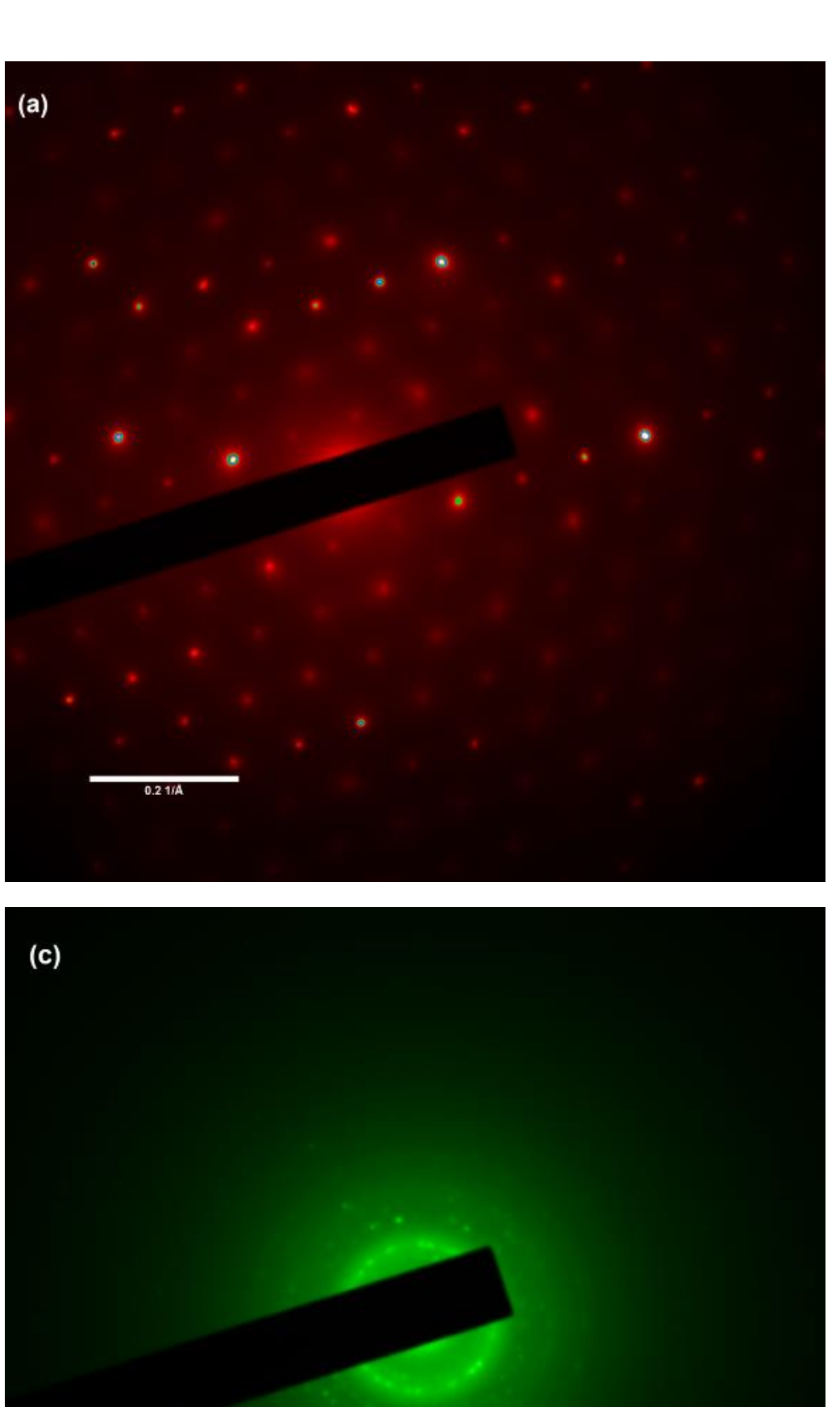

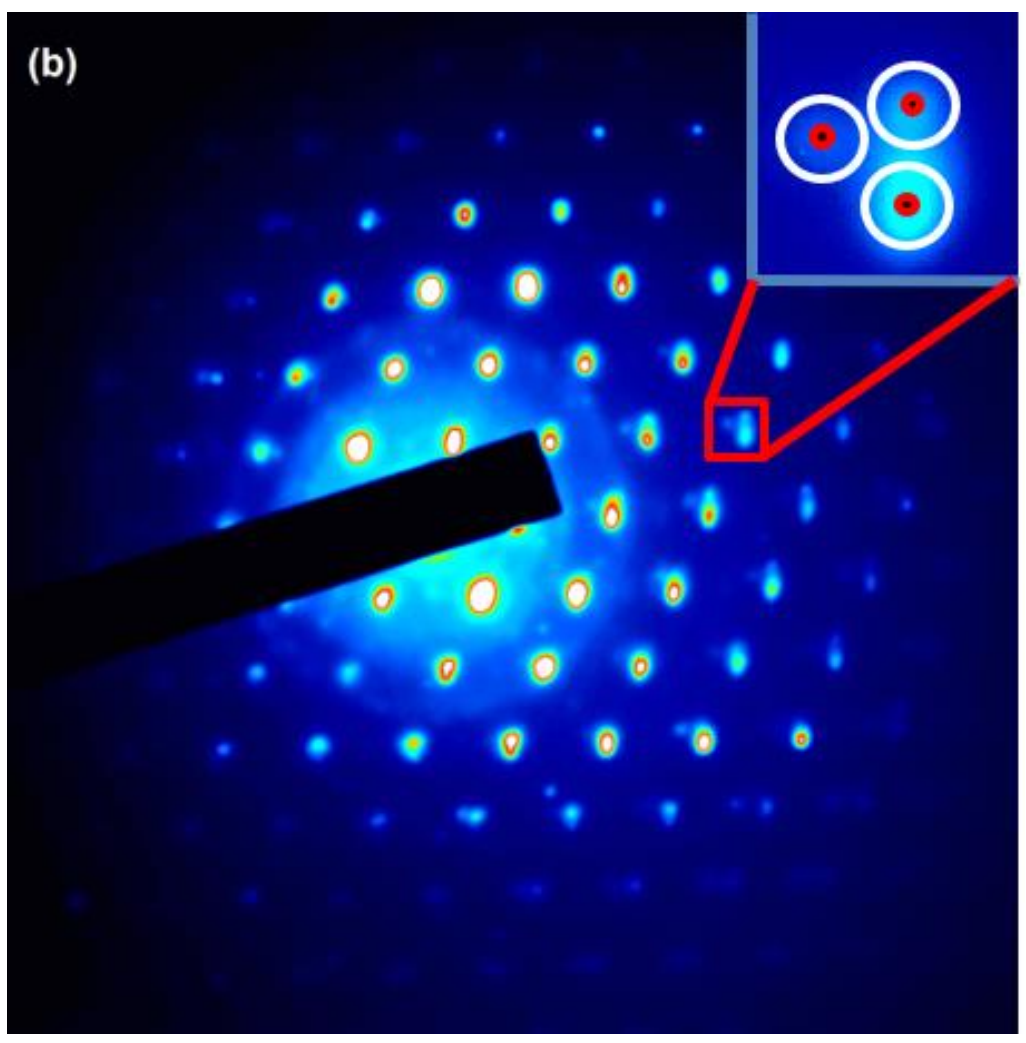

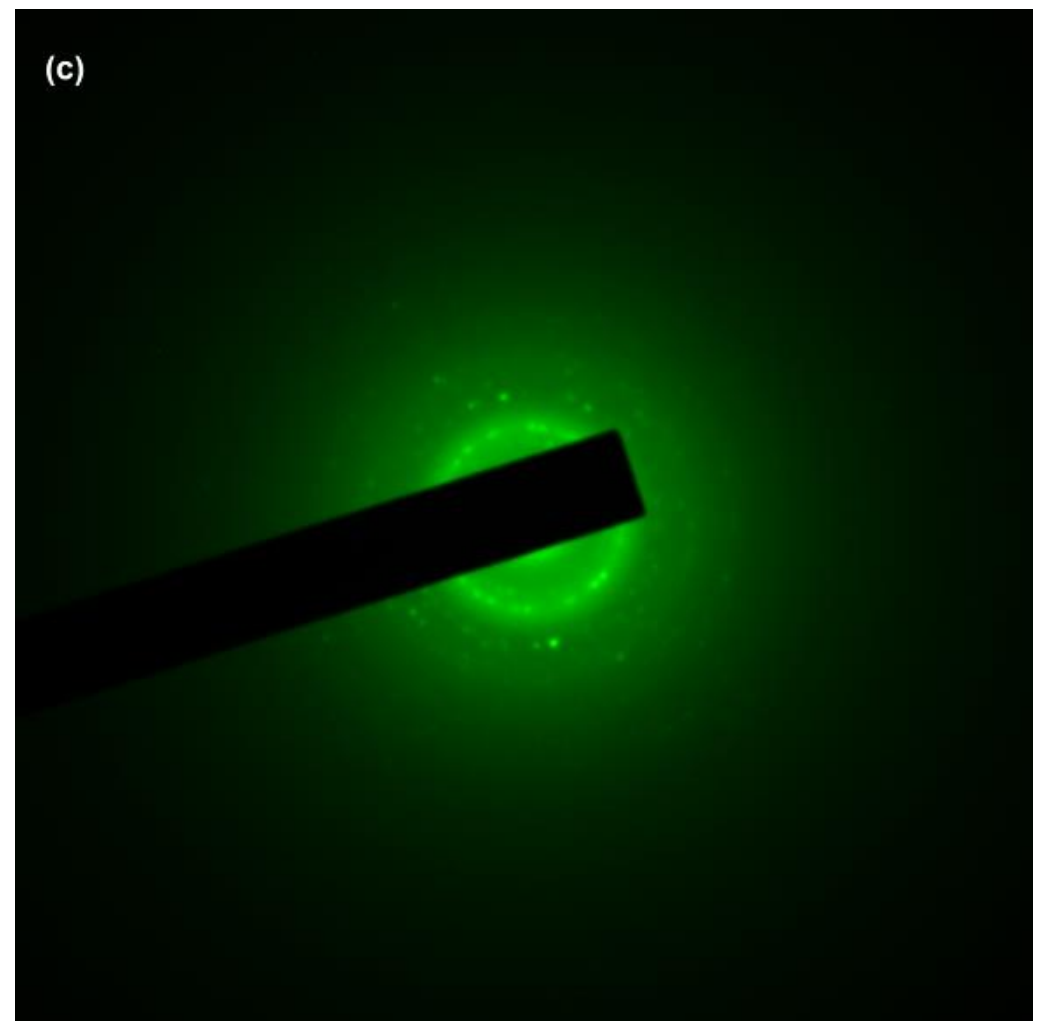

**Fig. 3:** (a) SAED pattern showing ordered hexagonal structure of the sample, (b) SAED pattern after around 1-2 min of e-beam exposure (inset shows three distinct lobes present in a diffraction spot), and (c) The same crystal showing nanocrystallinity after > 2 min of exposure.

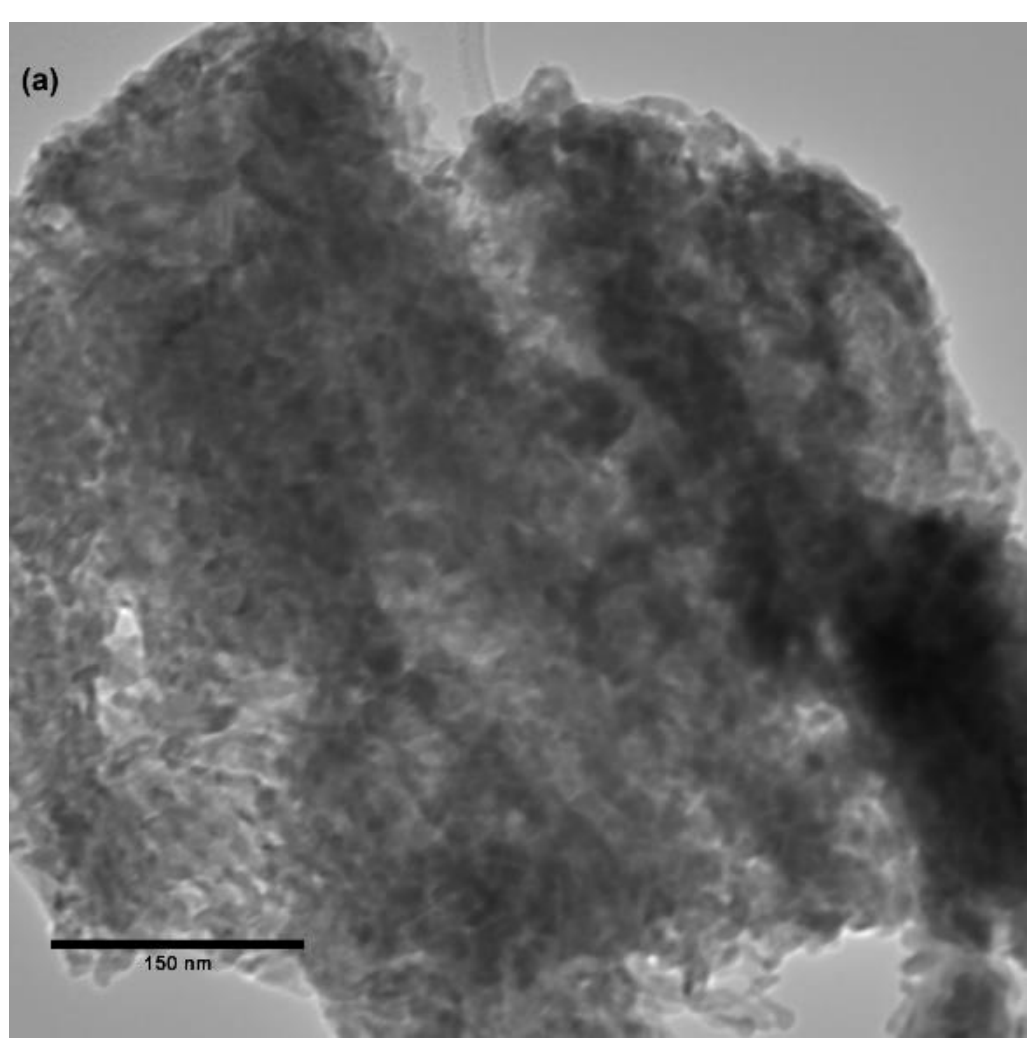

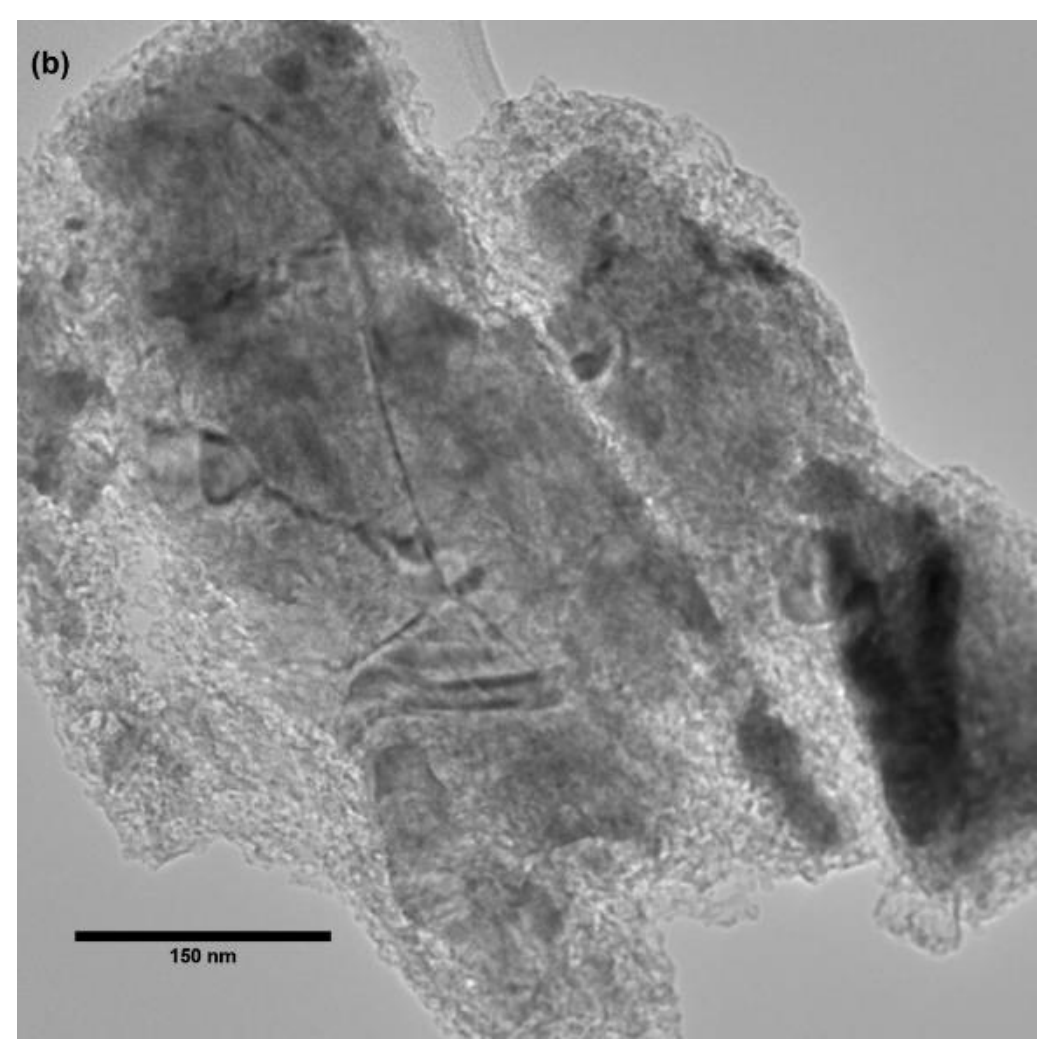

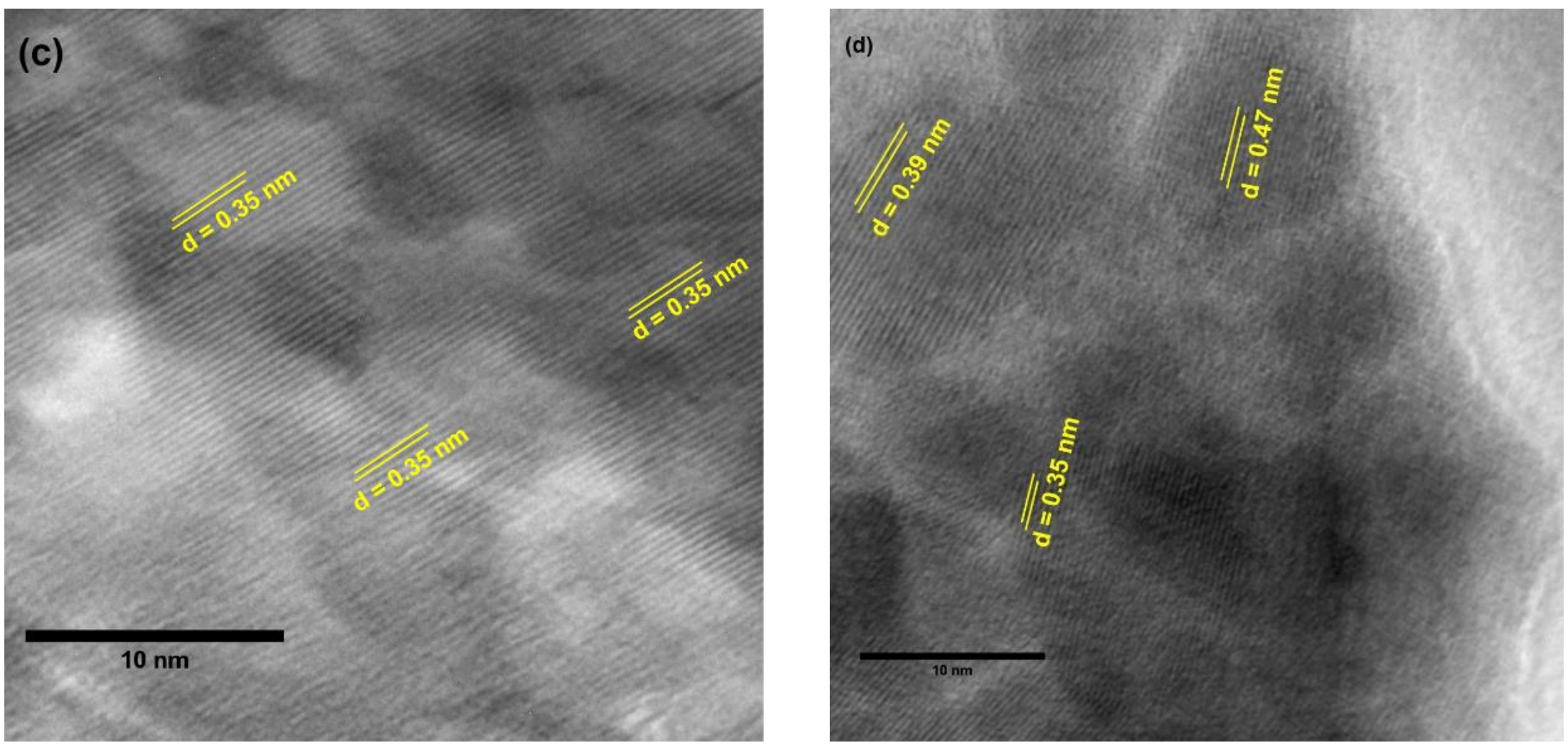

**Fig. 4:** (a), (b) TEM images of the sample taken before and after exposure to electron beam. (c) HR-TEM image showing uniform lattice fringes, and (d) HR-TEM image after domain formation showing changes in the lattice fringes.

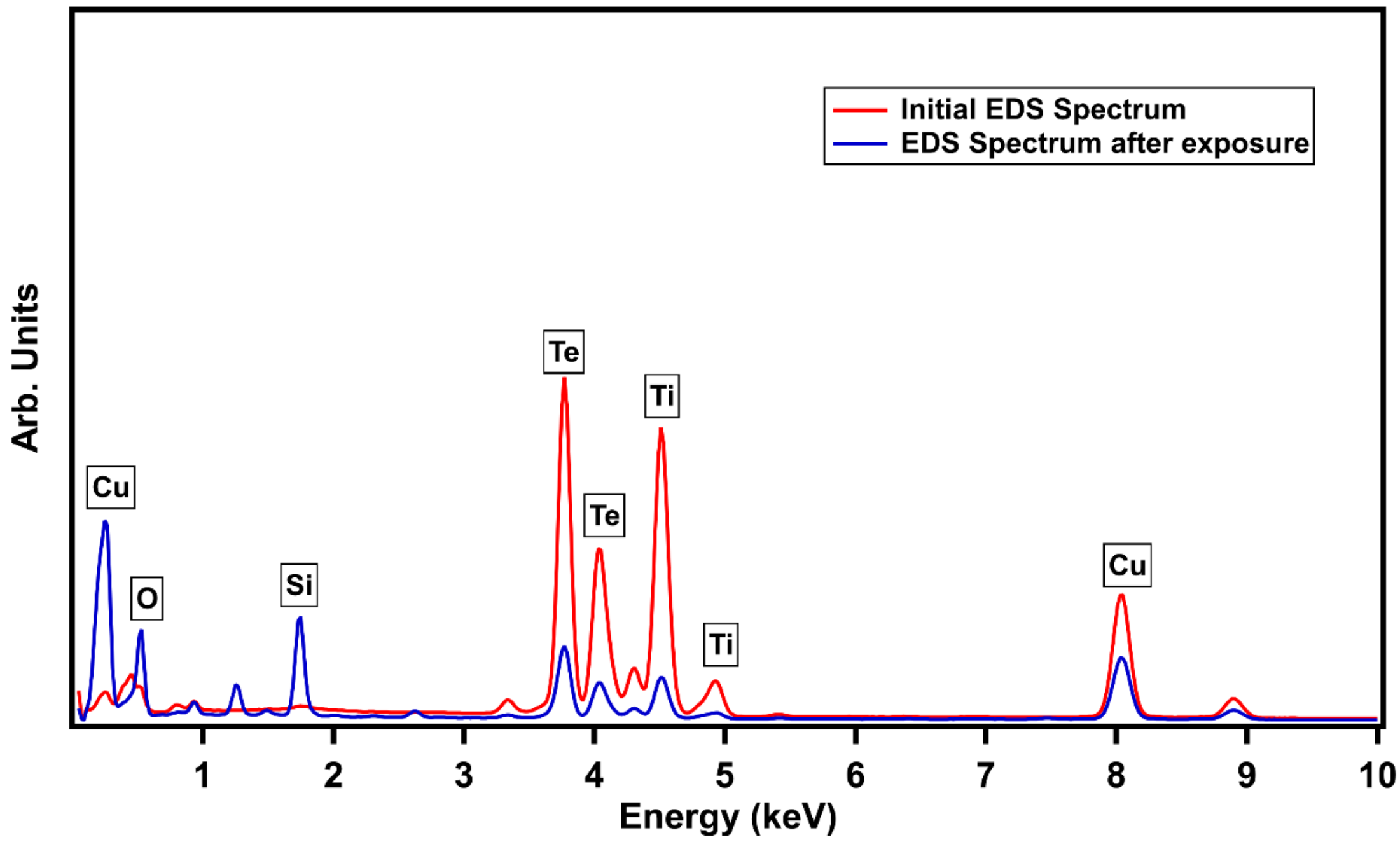

**Fig. 5:** Elemental analysis of the sample using EDS: red line (before) and blue line (after significant exposure to e-beam).

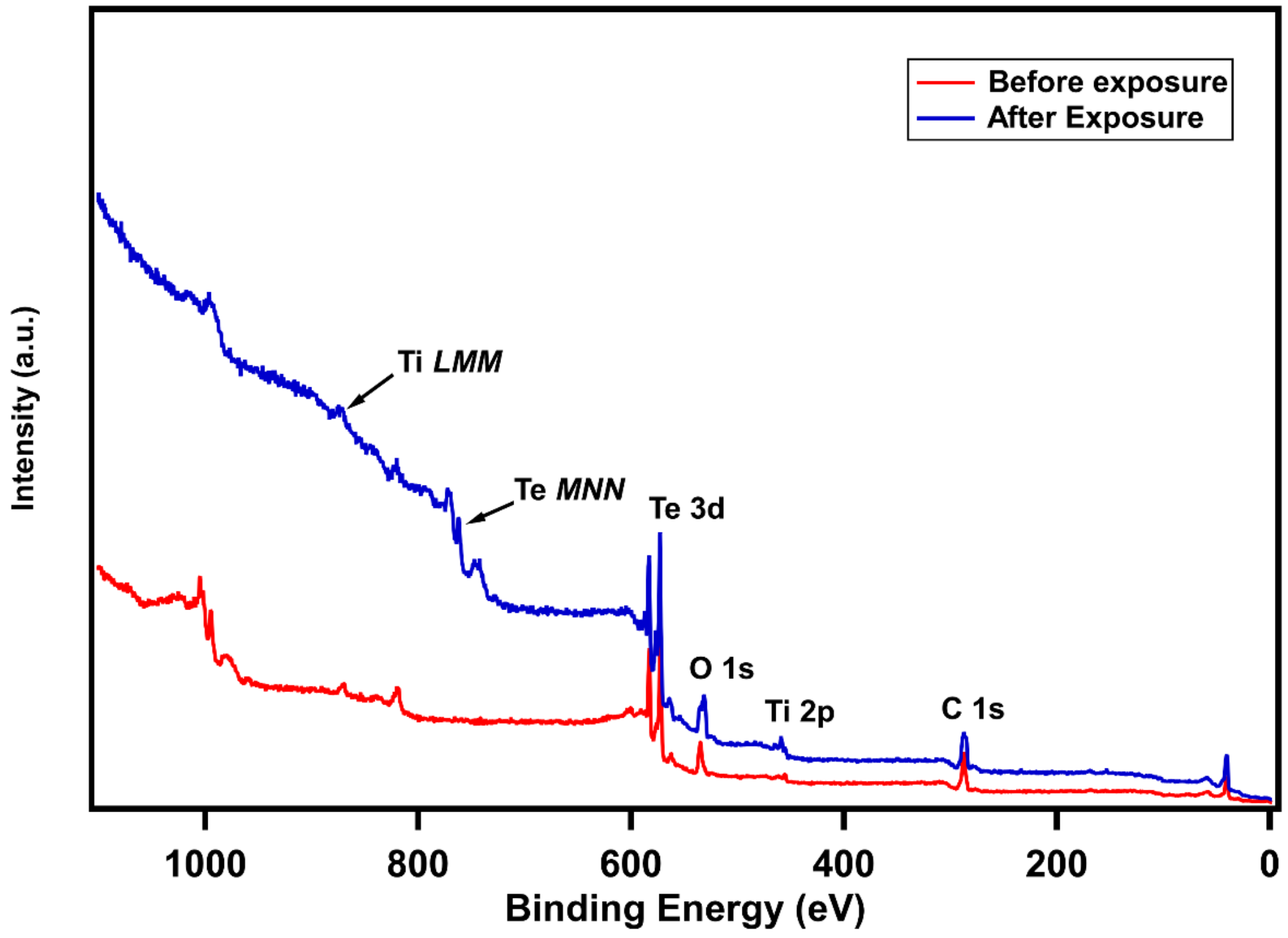

**Fig. 6:** XPS survey scan of the sample denoting the characteristic core levels and Auger lines at higher binding energies after exposure to air.

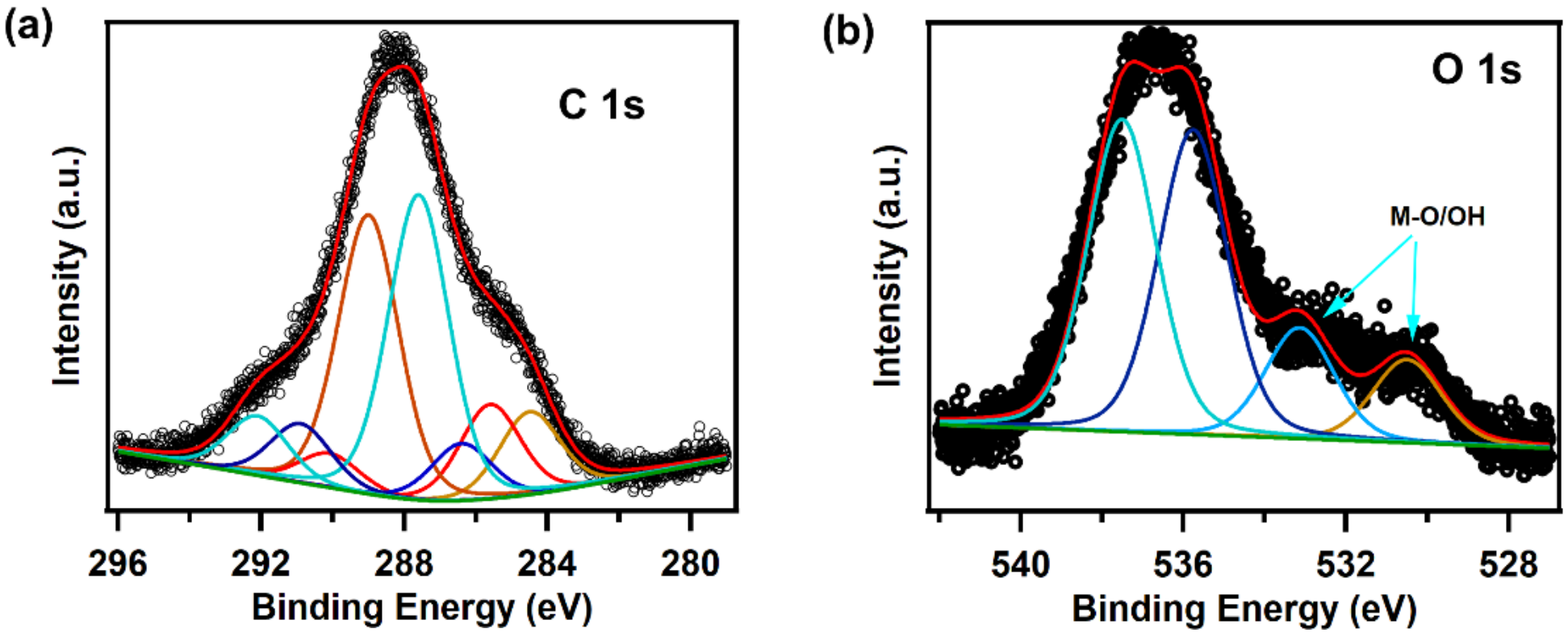

**Fig. 7:** XPS core level spectra for (a) C1s and (b) O 1s. O 1s shows presence of metal oxides and hydroxides.

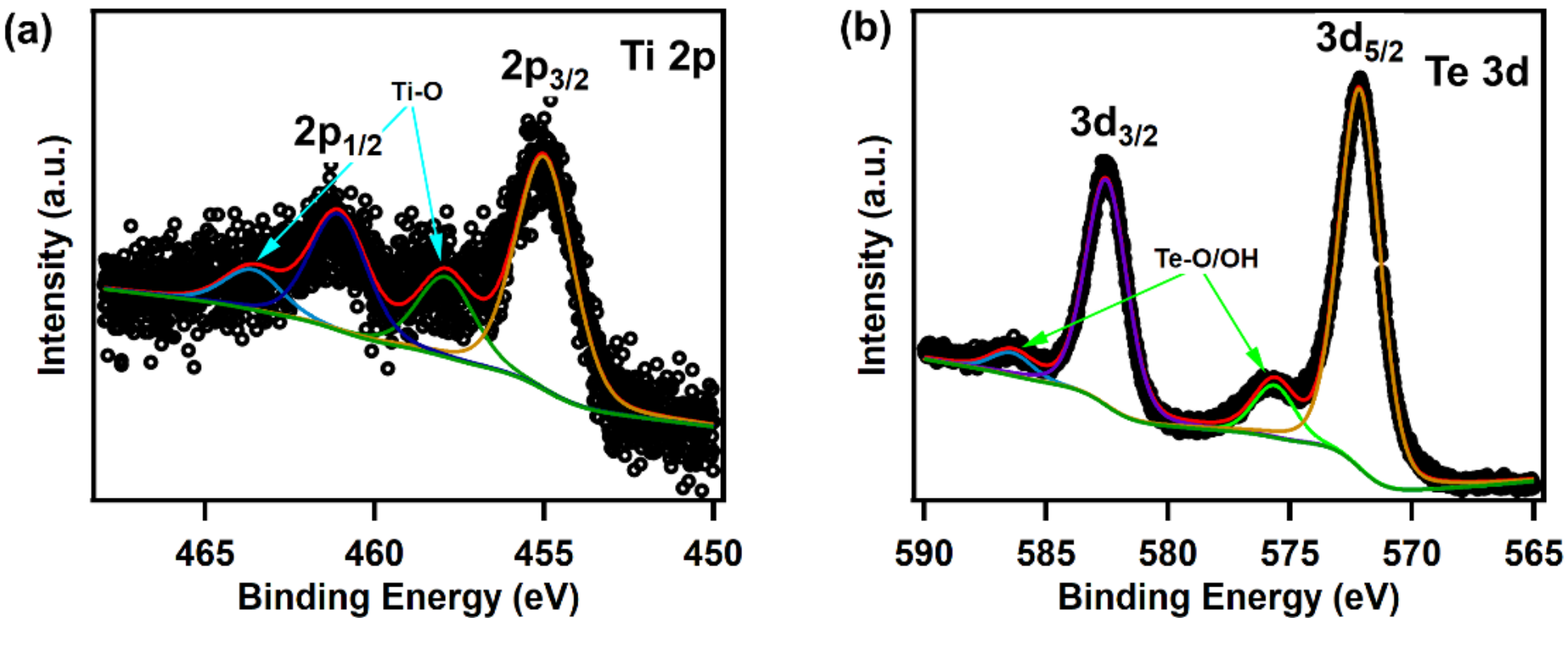

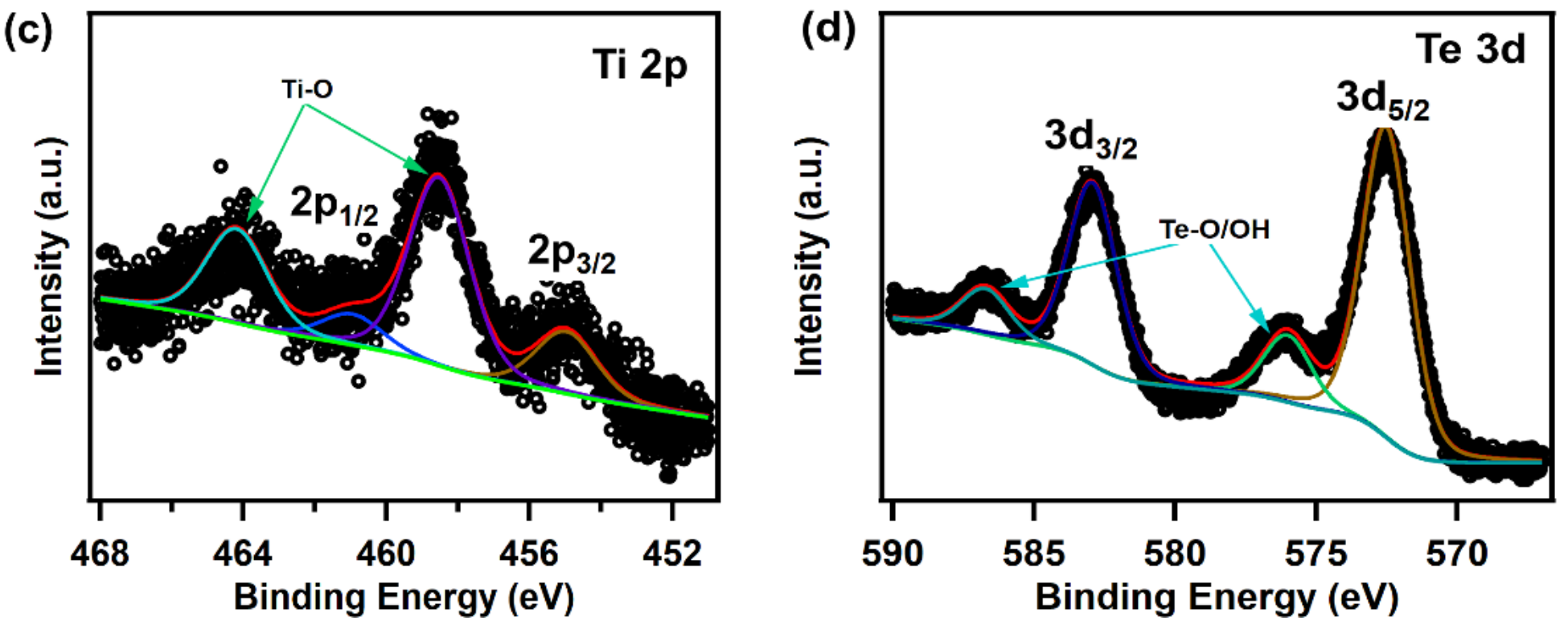

**Fig. 8:** Comparison of XPS core level spectra of Ti 2p and Te 3d. Panels (a), (b) correspond to initial spectra and (c), (d) to the spectra acquired after significant air exposure.

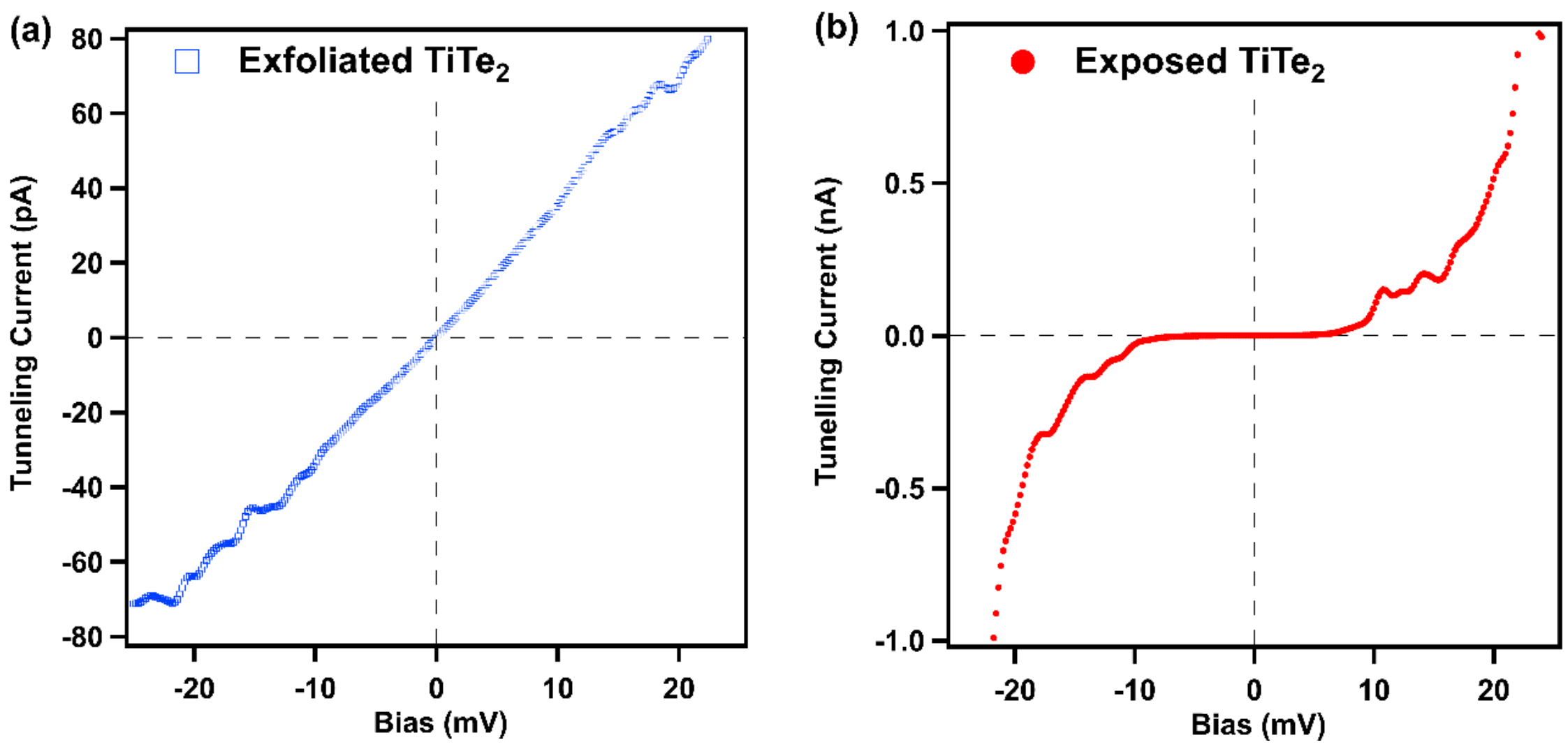

**Fig. 9:** I-V curve of freshly exfoliated sample showing ohmic behavior and air exposed sample showing semiconducting pattern.